# What do student responses to the car-truck problems tell us? An investigation into two Force Concept Inventory questions

Ashutosh Kumar Pathak[1*], Mark A. J. Parker[1], Ross K. Galloway[2], Andrew J. A. James[1], Sally E. Jordan[1], Jonathan Nylk[1]

1. School of Physical Sciences, The Open University, Milton Keynes, MK7 6AA, UK
2. School of Physics and Astronomy, University of Edinburgh, Edinburgh, EH9 3FD, UK

*Contact author: ashutosh.pathak@open.ac.uk

**Abstract**

Concept inventories are widely used in Physics Education Research to uncover student misconceptions and measure the effectiveness of instruction. Two questions in the widely used multiple-choice Force Concept Inventory concern a car that is pushing a truck, as the car is accelerating and then when a steady speed has been reached. We investigate the premise that correct answering of these questions is because of good conceptual understanding of Newton's Third Law, by analyzing responses to a modified inventory that randomly offers students multiple-choice or free-text versions of the questions and follows these with questions in which students select the physical principle(s) they used. The results from 674 attempts from students at six universities show a similar pattern of results for multiple-choice and free-text versions, also comparable with Force Concept Inventory results from one of the universities in the preceding five years, but with marked differences relative to early published Force Concept Inventory data. Free-text responses and the law-selection sub-questions can reveal more about conceptual understanding than can be seen in multiple-choice responses alone. For the accelerating system (Question 15), correct responses are usually attributed to Newton's Third Law and incorrect responses are usually attributed to Newton's Second Law. When speed is constant (Question 16), around 90% of responses are correct, but a significant number of these are attributed to Newton's First Law or the Superposition Principle rather than Newton's Third Law, especially pre-instruction. We conclude that students are confusing balanced forces on a single object moving at constant speed with the correct equal-and-opposite interaction pair between two objects, and correct responses to Question 16 may be concealing an underlying misconception.

## 1. INTRODUCTION

Since its introduction in 1992, the Force Concept Inventory (FCI) [1] has been widely used to assess students' conceptual understanding of Newtonian mechanics. The original version of the FCI consisted of 29 multiple-choice items. Minor revisions in 1995 led to a 30-item inventory [2], which is the version that is still commonly used, and this version forms the reference point for the work described here. The FCI was designed to associate each question with a particular Newtonian law, while each incorrect option for the multiple-choice instrument was aligned with a particular misconception. Thus, by design, the instrument provides educators with information about their students' conceptual understanding and, when administered before and after instruction, gives a measure of the learning that has taken place, indicating the effectiveness of the instruction [3].

It is generally agreed that performance on the FCI correlates with conceptual understanding, but for many years there has been debate regarding the detail of what is actually measured [4, 5]. Countless educational evaluations have been conducted into the use of the instrument, with many focusing on gender and ethnic disparities in performance [e.g. 6-8], though opinion remains divided as the underlying causes of these. The multiple-choice (MC) format is one suggested cause of the demographic disparities [9], though again consistent evidence is absent from the literature. More generally, while the MC format makes the instrument easy to administer and mark, there is a concern that the ability to guess an answer might mean that students appear to have good conceptual understanding when this is not the case [10]. There is also concern that, even when a student has not obtained the answer entirely by guesswork, other aspects of the question or the options provided may have guided them to the correct answer [11] or misled them [12]. Crucially, in its current form, the FCI cannot provide any information about why a particular option was selected [13] and it has been shown that students can often arrive at a correct answer through flawed reasoning [14], creating a "false-positive", where the response leads to an unjustified assumption about the level of conceptual understanding [15].

A possible alternative to relying on MC questions is to instead use free-text (FT) questions, in which students compose their own responses rather than selecting from options provided. It has been suggested that FT assignments can provide insight beyond that obtained from conventional MC concept inventories [16, 17]. Student responses to free-text questions have always had an important role in the development of concept inventories, by suggesting possible distractors for inclusion in a later MC instrument [18]. However, the rapid 21st Century growth in the availability and accuracy of automatic marking of FT responses [19, 20] has opened up the possibilities for the use of FT questions as an alternative to MC in the concept inventories themselves [21-23].

Similarly, sub-questions, usually additional concept inventory type questions, designed to assess the same concept but in a different context, have been used to test the validity of FCI items [24], in particular as a way of reducing false positives [15]. Questions that ask students to give the reasoning behind their answer can be incorporated into a FCI question [23] or added as sub-question.

The present study forms part of a wider investigation into the use of FT questions and sub-questions in a modified FCI. Here we concentrate on two FCI questions which concern a car that is pushing a truck, firstly as the car is accelerating (Question 15 in the commonly used version of the FCI [2]) and secondly when a steady speed has been reached (Question 16). Correct answering of these questions is generally attributed to good understanding of Newton's Third Law, but concern has been raised as to whether this is actually the case [25]. The incorrect options (known as distractors) provided in the conventional MC version of the FCI are identical for Q15 and Q16 and their selection is attributed to the following misconceptions [1]:

Option B: greater mass implies greater force

Option C: most active agent produces greatest force

Option D: only active agents produce force

Option E: obstacles exert no force

These two questions were offered to students in either MC or FT format, and each was followed by a sub-question asking students to identify which of a number of specified physical laws they had used in reaching their answer.

The research questions that we were seeking to investigate were:

- How does the distribution of responses to the target questions from students in the study compare with those found by others?
- Do equivalent MC and FT versions of the same question lead to the same conclusions about students' conceptual understanding?
- Do students correctly apply Newtonian laws in answering the questions?

## II. METHODS

### A. Instrument design and development

The questions of particular interest in the present study were included in a 38-item test referred to as the Newtonian Mechanics Quiz, containing a mixture of FCI questions (including the 14 items from the one of two "half FCIs", which have been found to produce scores that are highly correlated with those for the full FCI [26]) and new sub-questions.

Five of the FCI questions, including two of those discussed in the present study, were randomly offered to students in conventional MC format or in FT format, requiring students to type a free-text response of up to 20 words. The two formats of the first question of interest are compared in Fig. 1. The FT versions of these questions had been refined and evaluated as part of an earlier project [27]. In the current study, student responses were collected online but marked manually.

**Use the information below to answer questions 34 to 37:**

A large truck breaks down out on the road and receives a push back into town by a small compact car as shown in the figure below.

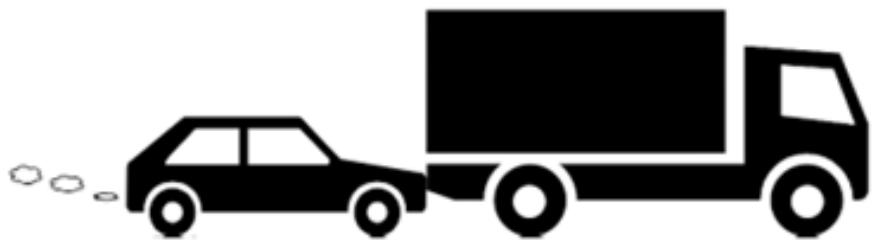

**MC version of Q34**
While the car, still pushing the truck, is speeding up to get up to cruising speed,

A. the amount of force with which the car pushes on the truck is equal to that with which the truck pushes back on the car.
B. the amount of force with which the car pushes on the truck is smaller than that with which the truck pushes back on the car.
C. the amount of force with which the car pushes on the truck is greater than that with which the truck pushes back on the car.
D. the car's engine is running so the car pushes against the truck, but the truck's engine is not running so the truck cannot push back against the car. The truck is pushed forward simply because it is in the way of the car.
E. neither the car nor the truck exerts any force on the other. The truck is pushed forward simply because it is in the way of the car.

**FT version of Q34**
While the car, pushing the truck, is speeding up, how does the force that the car exerts on the truck compare with the force that the truck exerts on the car?

FIG. 1. Comparison of MC and FT versions of Q15 (NMQ Question 34).

The questions from the FCI were supplemented by two types of sub-questions:

(i) Additional concept-inventory type questions, specially written to complement or follow-up on a particular FCI question and so gain additional insight into conceptual understanding. None of these questions are included in the current study.

(ii) Questions requiring students to specify which of a number of physical law(s) they used in answering the preceding FCI question. The same wording (shown in Fig. 2) was used on each occasion, including for two questions in the current study. These questions used a multiple-response format, enabling students to select as many options as they felt to be appropriate.

37. Which of the following physical law(s) did you use to support your response in question 36? Tick all that apply.

A. Newton's first law: A body remains at rest or in a state of uniform motion unless it is acted on by a resultant force.
B. Newton's second law: A resultant force acting on a body of fixed mass will cause to accelerate in the direction of the resultant force.
C. Newton's third law: For every action there is always an equal and opposite reaction: or, the mutual actions of two bodies upon each other are always equal and in opposite directions.
D. Superposition principle: Two or more forces of the same magnitude but opposite direction act on an object, their vector sum will be zero, resulting in a state of equilibrium.

FIG. 2. The law-selection sub-question used after various NMQ questions, shown in its use as Q16LS (NMQ Question 37).

For ease of comparison, we refer to the questions that are identical to or the FT versions of Question 15 in the commonly used version of the FCI [2], considering an accelerating car pushing a truck, as Q15, even though this question had a different number in the NMQ. Similarly, questions that are identical or FT versions of FCI Question 16, considering a car moving at constant speed and still pushing a truck, are always referred to as Q16. The law-selection sub-questions following Q15 and Q16 are referred to as Q15LS and Q16LS respectively.

The NMQ went through an iterative design process prior to the work described in the present article, incorporating feedback from 12 university professors and pilot trials and interviews with a group of university students. Further details about the development and evaluation of the NMQ are given elsewhere [28].

### B. Data collection and analysis

After development, the NMQ was offered to first year physics students from six different universities in the UK during the academic year 2024 to 2025, and our complete dataset of responses is based on 674 student attempts. The work described in this study is based primarily on 277 pre-instruction and 245 post-instruction (hereafter abbreviated "pre-test" and "post-test") attempts from University A, a Scottish research-intensive University.

To place the results in context, pre-test and post-test student responses to Q15 and Q16 in the conventional FCI, gathered from students at University A over the preceding five academic years (2019/20 to 2023/24) were first analyzed. These results were compared with both the data reported by Hestenes et al. [1] from analysis of early FCI use and with the results of the main 2024/25 study based on the NMQ.

To most fairly compare the performance of the MC and FT versions of the questions, both before and after instruction, the initial analysis considered only the subset of responses from students who had received a question in the same format pre-test and post-test. Since the allocation of MC or FT versions of each instance of each question was completely random, this reduced the dataset to 45 MC responses and 47 FT responses to Q15 and to 42 MC responses and 57 FT responses to Q16. This small matched dataset brought greater insight into learning gain and meant that the students who did not attempt the post-quiz, likely to be weaker and thus to skew the results, were excluded. However, because of the small size of the matched dataset, the analysis was repeated for the entire University A dataset in order to get a more complete picture.

Before FT responses could be analyzed, each was marked as correct or incorrect and also classified as corresponding to MC options A to E, or “X”, a category used for responses which were not considered to be sufficiently close to one of the MQ options. The FT responses were independently marked by four of the authors and where there was initial disagreement, the responses were discussed, with the first-named author making the final decision in borderline cases. For consistency, the first-named author was responsible for the classification of the incorrect responses, recognizing that this was necessarily subjective in some cases. The precise wording of each FT response was inspected to check for hints as to the student's reason for giving the answer that they did.

Student responses to the law-selection questions Q15LS and Q16LS were also analyzed, alongside their responses to Q15 and Q16 respectively, to ascertain whether students who selected the correct response also identified the Newtonian law(s) relevant to the situation, and vice versa. Inspection of the pattern of responses received across all four questions prompted further quantitative further analysis into patterns and correlations, in particular between a student's response to Q16LS and their other answers.

To verify the extent to which the findings for University A have applicability across the UK, the findings were then compared with results from the remaining 112 attempts, all gathered pre-instruction from students at five different Universities in England.

## II. RESULTS AND ANALYSIS

### A. Analysis of historic data

For University A, the results for both Q15 and Q16, given in detail in Appendix 1, appear to have been stable for the five years of the analysis so it was felt that combining them was justified, as done in Table I and Table II. The Hestenes results, also given in Table I and Table II.  varied considerably from cohort to cohort, unsurprisingly because some (AZR, AZH, MW and GS) were from high schools and some (AVH and HU) were from universities, though the mean overall FCI pre-test score at for the AVH group was found to be not much higher than that for the high schools, and other studies have found consistent learning gain across diverse groups [3]. The nature of instruction between pre- and post-test varied considerably which has been found to have a significant impact on learning gain [3]. These factors meant that it was not felt that the datasets could be combined which means that we do not have a large dataset from the time of the FCI's development against which to compare more recent results. The results should be interpretated with caution because of the small size of some of the cohorts. However, there are some similarities and differences in the data presented in Tables I and II that are worth commenting on.

Student responses to Question 16 were consistently more likely to be correct than those to Question 15.  For Question 15 (Table 1), Option C was consistently the most popular choice pre-test, with a decrease in selection of this option post-test and an increase in the selection of Option A (the correct option). For University A the percentage of responses selecting Option C decreased from 56% (pre-test) to 27% (post-test) while the percentage of responses selecting Option A increased from 40% (pre-test) to 71% (post-test). The pre-test responses in the earlier dataset were noticeably less likely to be correct, with the percentage selecting Option A ranging from 6% to 14%. However, there were significant improvements by the time of the post-test for some of the groups, notably GS and MS which both benefitted from innovative pedagogy, with 78% or 79% now selecting Option. A. At University A, the selection of Options B, D and especially E was very uncommon. For some institutions in the earlier data set, Option B and Option D were more popular, especially pre-test, though Option E was unpopular then too.

In response to Question 16 (Table II), 91% of students at University A gave Option A (the correct option) pre-test. This left limited room for improvement post-test as a result of the so-called ceiling effect[29], and 95% of students selected Option A post-test. Hestenes et al. report rather less successful pre-test results with 35% to 56% of the cohort selecting Option A, with an increase from pre-test to post-test of 29 to 52 percentage points. Only post-test results are presented for HU but more than 90% of students in this cohort and some of the others selected Option A post-test. For almost all the groups, both pre-test and post-test, Option C was most popular incorrect option.

Table I. Results for University A compared with those in [1] for the question that is now FCI Q15. The cohorts and the institutions they were from are identified in [1].

| **Institution** | **Pre- or Post-test** | **Number of students** | **Percentage of students who chose option** | | | | |
|---|---|---|---|---|---|---|---|
| | | | **A** | **B** | **C** | **D** | **E** |
| University A (2019-24 combined) | Pre-test | 1098 | 40% | 2% | 56% | 1% | <1% |
| | Post-test | 727 | 71% | 1% | 27% | <1% | 0% |
| AZR | Pre-test | 612 | 8% | 17% | 62% | 12% | 1% |
| | Post-test | | 19% | 9% | 69% | 2% | 0% |
| AZH | Pre-test | 118 | 6% | 9% | 75% | 10% | 0% |
| | Post-test | | 23% | 7% | 70% | 0% | 0% |
| MS | Pre-test | 30 | 14% | 6% | 69% | 11% | 0% |
| | Post-test | | 79% | 0% | 21% | 0% | 0% |
| GS | Pre-test | 63 | 8% | 13% | 75% | 5% | 0% |
| | Post-test | | 78% | 0% | 22% | 0% | 0% |
| AVH | Pre-test | 116 | 7% | 13% | 74% | 4% | 1% |
| | Post-test | | 47% | 2% | 51% | 0% | 0% |
| HU | Post-test | 186 | 57% | 4% | 39% | 0% | 0% |

Table II. Results for University A compared with those in [1] for the question that is now FCI Q16. The cohorts and the institutions they were from are identified in [1].

| **Institution** | **Pre- or Post-test** | **Number of students** | **Percentage of students who chose option** | | | | |
|---|---|---|---|---|---|---|---|
| | | | **A** | **B** | **C** | **D** | **E** |
| University A (2020-24 combined) | Pre-test | 810 | 91% | <1% | 5% | <1% | 2% |
| | Post-test | 526 | 95% | <1% | 2% | <1% | 1% |
| AZR | Pre-test | 612 | 35% | 9% | 40% | 13% | 3% |
| | Post-test | | 64% | 3% | 27% | 2% | 3% |
| AZH | Pre-test | 118 | 39% | 5% | 43% | 12% | 1% |
| | Post-test | | 68% | 1% | 28% | 3% | 1% |
| MW | Pre-test | 30 | 56% | 3% | 28% | 11% | 3% |
| | Post-test | | 93% | 0% | 3% | 0% | 3% |
| GS | Pre-test | 63 | 30% | 6% | 53% | 5% | 6% |
| | Post-test | | 75% | 0% | 11% | 0% | 14% |
| AVH | Pre-test | 116 | 37% | 4% | 49% | 6% | 4% |
| | Post-test | | 89% | 0% | 11% | 0% | 0% |
| HU | Post-test | 186 | 91% | 1% | 3% | 0% | 5% |

## B. Comparison of responses from multiple-choice and free-text versions of questions

### *1. Q15*

Table III compares the performance of the students from the small matched dataset of University A students who received either the MC or FT version of Q15 at both pre-test and post-test. The able shows that the most common pre-test response for both formats was Option C, while the most common post-test response was Option A (the correct option).

TABLE III. Comparison of options identified for the MC and FC versions of Q15 in pre- and post-test for the matched dataset from University A.

| Pre- or post-test | MC or FT | Size of group | Number of responses aligned with each option | | | | | |
|---|---|---|---|---|---|---|---|---|
| | | | A | B | C | D | E | X |
| Pre-test | MC | 45 | 21 | 1 | 23 | 0 | 0 | - |
| | FT | 47 | 16 | 0 | 26 | 0 | 0 | 5 |
| Post-test | MC | 45 | 30 | 2 | 13 | 0 | 0 | - |
| | FT | 47 | 33 | 1 | 11 | 0 | 0 | 2 |

The same trend was obtained when all of the MC and FT responses for University A were considered (Fig. 3). The results were also broadly in line with those reported for University A in previous years (Table I).

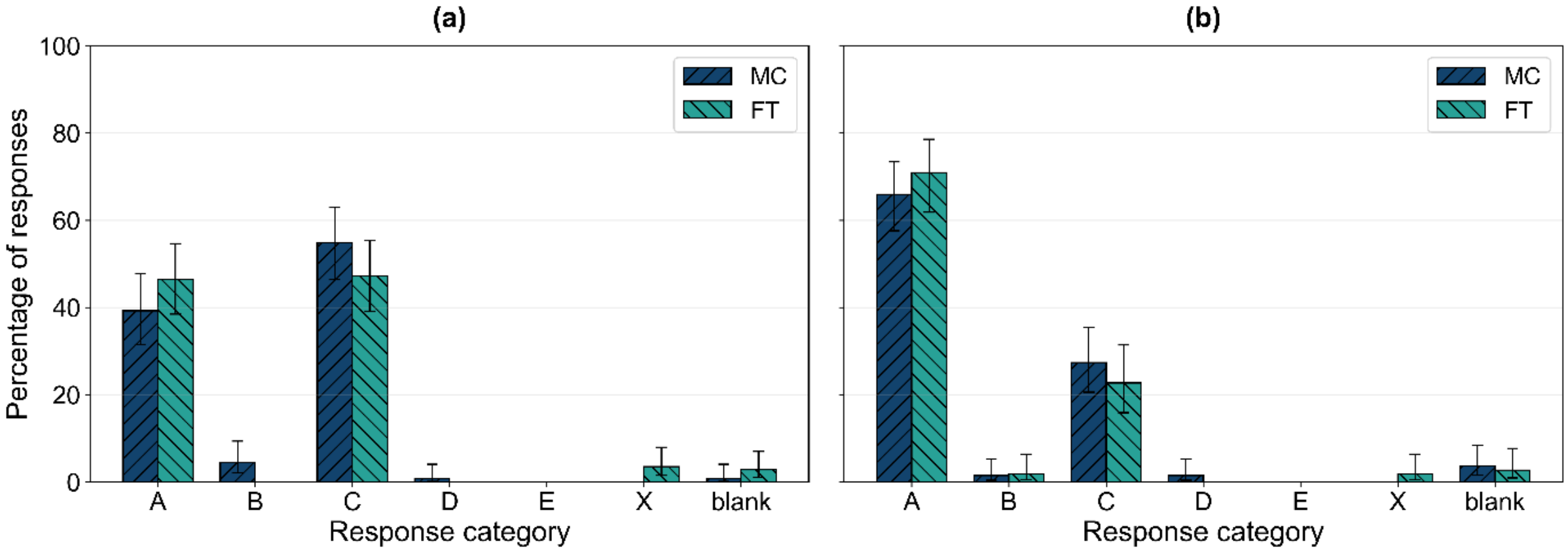

FIG. 3. Comparison of MC and FT responses to Q15 for the full University A dataset (a) pre-test (135 MC and 142 FT responses) and (b) post-test (135 MC and 110 FT responses). The error bars indicate 95% confidence intervals.

There were a small number of FT responses that were not identified as aligning to any of the MC options. These were classified as “X” in Table III and Fig. 3 and are listed below:

- Both forces increase
- They increase at the same rate
- The magnitude of the forces both increase at the same rate
- The force that the car exerts on the truck increases
- The car's force increases while the trucks does not, at first.
- They both increase, but truck force is larger because of F=ma
- the force exerted by the car increases as the force exerted by the truck remains constant

The first three of the listed responses are interesting because they all talk about the forces increasing, but they fail to compare the force that the car exerts on the truck with the force that the truck exerts on the car. These responses do not answer the question, which asked students to compare the force that the car exerts on the truck with the force that the truck exerts on the car, so they are incorrect. However, especially when a response talks about the forces increasing at the same rate, it seems quite likely that the student had a correct understanding of the physics which the question seeks to assess.

Many FT responses to Q15 that were considered to align with the MC options were very brief and, even when slightly longer, they did not give substantially more information than their MC equivalent as to the student's conceptual understanding. For example, "The forces are equal" and simply "equal" or "same" were all classified as A while "The car exerts a greater force on the car than the truck exerts on the car" was classified as C, but these responses did not, by themselves, give more information as to the student's reason for giving their answer. The response "Equal but opposite" gave more information than a correct response to the MC version, where Option A does not mention the fact that the forces act in opposite directions. Also, some responses gave an indication as to the underlying conceptual understanding, either directly or indirectly, as shown in the following examples:

- They are equal (Newton's third law)
- Car's force is greater, truck remains at rest until acted on which causes truck to accelerate in same direction.
- force that the car exerts is greater since there is a acceleration, while force that truck exerts is smaller
- It is greater as the truck will accelerate in the direction of the cars unbalanced force acting on it
- The car exerts a greater force on the truck to allow it to move forward

The first example given above states explicitly that the student was using Newton's third law. Most of the other responses, which were all classified as aligning to Option C, mention acceleration, while the final one gives an indication that the student believed that the active agent produced the greater force.

*2. Q16*

Table IV, which compares the performance of the students from the small matched dataset who received the MC and FT versions of Q16, shows that almost all of the students gave Option A (the correct option) in the pre-test, irrespective of the format of the question. This apparently left only a small margin for further improvement (the ceiling effect), with all but one of the 42 students who received the MC version selecting Option A in the post-test, and all but one of the 57 students who received the FT version gave a response that was classified as equivalent to Option A.

TABLE IV. Comparison of options identified for the MC and FC versions of Q16 in pre- and post-test for the matched dataset from University A.

| **Pre- or post-test** | **MC or FT** | **Size of group** | **Number of responses aligned with each option** | | | | | |
|---|---|---|---|---|---|---|---|---|
| | | | **A** | **B** | **C** | **D** | **E** | **X** |
| Pre-test | MC | 42 | 38 | 1 | 3 | 0 | 0 | - |
| | FT | 57 | 54 | 1 | 1 | 0 | 0 | 1 |
| Post-test | MC | 42 | 41 | 0 | 1 | 0 | 0 | - |
| | FT | 57 | 56 | 0 | 0 | 0 | 0 | 1 |

Similar results are reflected in the whole University A dataset, as shown in Fig 4. 87% or 88% of responses in the pre-test and 90% of responses (for both formats) in the post-test were Option A or a free-text response equivalent to it. These results are also in line with those for University A in previous years.

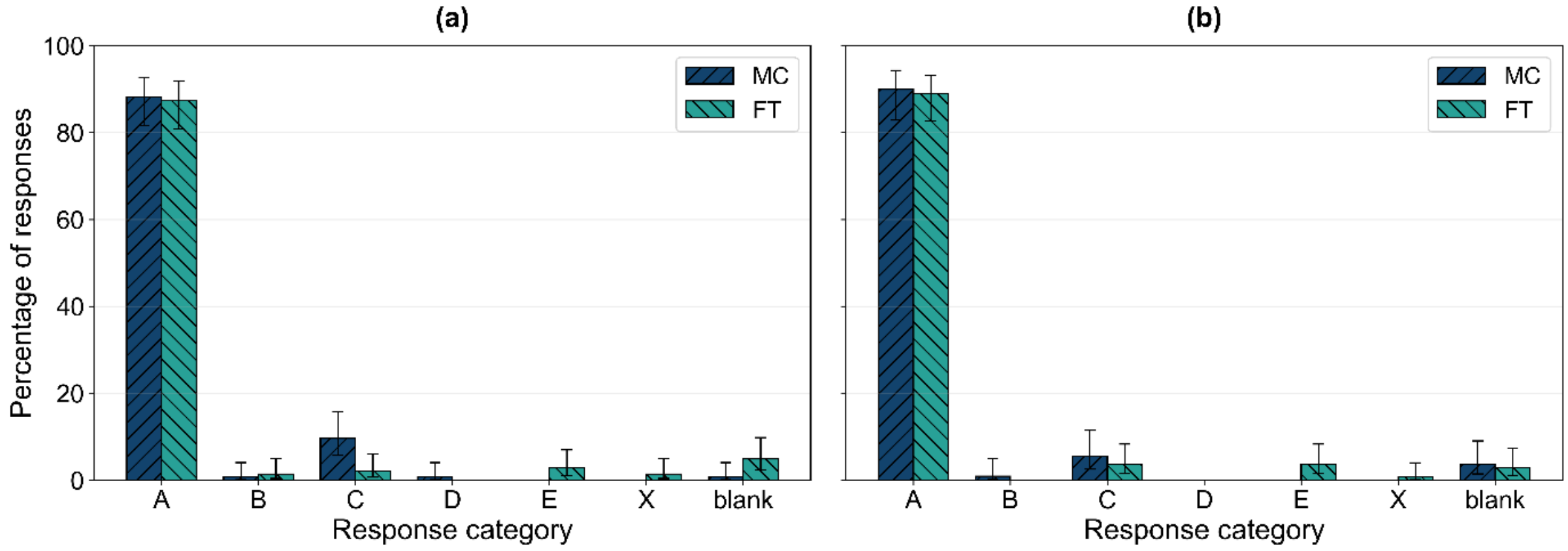

FIG. 4. Comparison of MC and FT responses to Q16 for the full University A dataset (a) pre-test (135 MC and 142 FT responses) and (b) post-test (109 MC and 136 FT responses). The error bars indicate 95% confidence intervals.

The FT responses to Q16 that were not identified as aligning to any of the MC options are listed below.

- Continue at same rate because truck is always pushing back
- Still remain constant
- They are in equilibrium, assuming the resistive forces on the car are included in the force of the car pushing.
- No force on the car
- They are both moving at the same speed so while they're in contact there is no force being applied
- There is no force between them as they are at a constant speed
- No resultant force
- The forces are equal which equal to zero
- The forces are equal to 0 as per newtons first law
- there is no force
- force is zero
- 0 force

Two groups of these responses are interesting. The first two responses both refer back to a response in Q15, illustrating the question-order effect[26], whereby the responses given reflect the order in which questions were asked [27]. However, perhaps the most interesting responses are those which said there was no force acting. It could be argued that these responses show partial alignment with Option E: "Neither the car nor the truck exerts any force on the other. The truck is pushed forward simply because it is in the way of the car." However, there is no indication that the students giving these responses believed the second part of Option E, namely that the truck is pushed forward simply because it is in the way of the car.

As for Q15, many Q16 responses that were classified as aligning to Option A were of the type "the same" or "equal" and it was not possible to tell from these alone, or even from some longer responses (e.g. "The forces acting on the car and the truck are equal and opposite"), whether the student had used correct or incorrect reasoning in reaching this conclusion. However, some responses did give an indication of the student's conceptual understanding, for example, where the responses were of the type "the forces are still the same" the student was arguably using the same reasoning as in Q15, suggesting the use of Newton's Third Law on both occasions. In contrast "the forces are now equal" indicated that the student had probably given an incorrect response to Q15, and most probably had some conceptual misunderstanding.

Where responses to Q16 talked about the forces being "balanced", this gave a slight indication that the student was confusing the scenario which applies in this question (mutual equal and opposite forces between two objects), with balanced forces acting on a single object when it is moving at constant speed. A few responses explicitly explained that the student had based their answer on the fact that the speed was constant, e.g. "Balanced going at constant speed", "they are still equal ('constant speed')".

### C. Responses to the law-selection sub-questions

It is interesting to note that a small number of students selected all of the available options in response to Q15LS and/or Q16LS. It is not clear whether these students thought that they had used all of Newton's First Law (N1), Newton's Second Law (N2), Newton's Third Law (N3), and the Superposition Principle (SP) in answering the preceding question or whether they were hoping that in selecting all options they would inevitably select the correct option(s) and so get some credit. In reality, for the purpose of this study, we did not grade the law selection sub-questions, but simply used them as an interesting indicator of student understanding.

#### *1. Q15LS*

For Q15LS, at both pre-test and post-test, a very different pattern of responses was observed dependent on whether the student's answer to Q15 was correct or incorrect, as shown as an UpSet plot [30] in Fig. 5. The horizontal bars shown below and to the left of the plot represent the total number of students who selected each law (irrespective of whether they had selected the law alone or in combination with others) while the vertical bars show the percentage of the population who gave each combination. Although less than half of the students gave a correct answer to Q15 at pre-test, 95% of the 119 who did also correctly selected N3 in response to Q15LS and most of those selected N3 alone, showing confident understanding. In comparison, for the 153 students who gave an incorrect answer to Q15 at pre-test, only 39% selected N3 in response to Q15LS while 88% selected N2. A higher percentage (69% overall) of students gave a correct answer to Q15 at post-test, and 95% of these selected N3 in response to Q15LS.

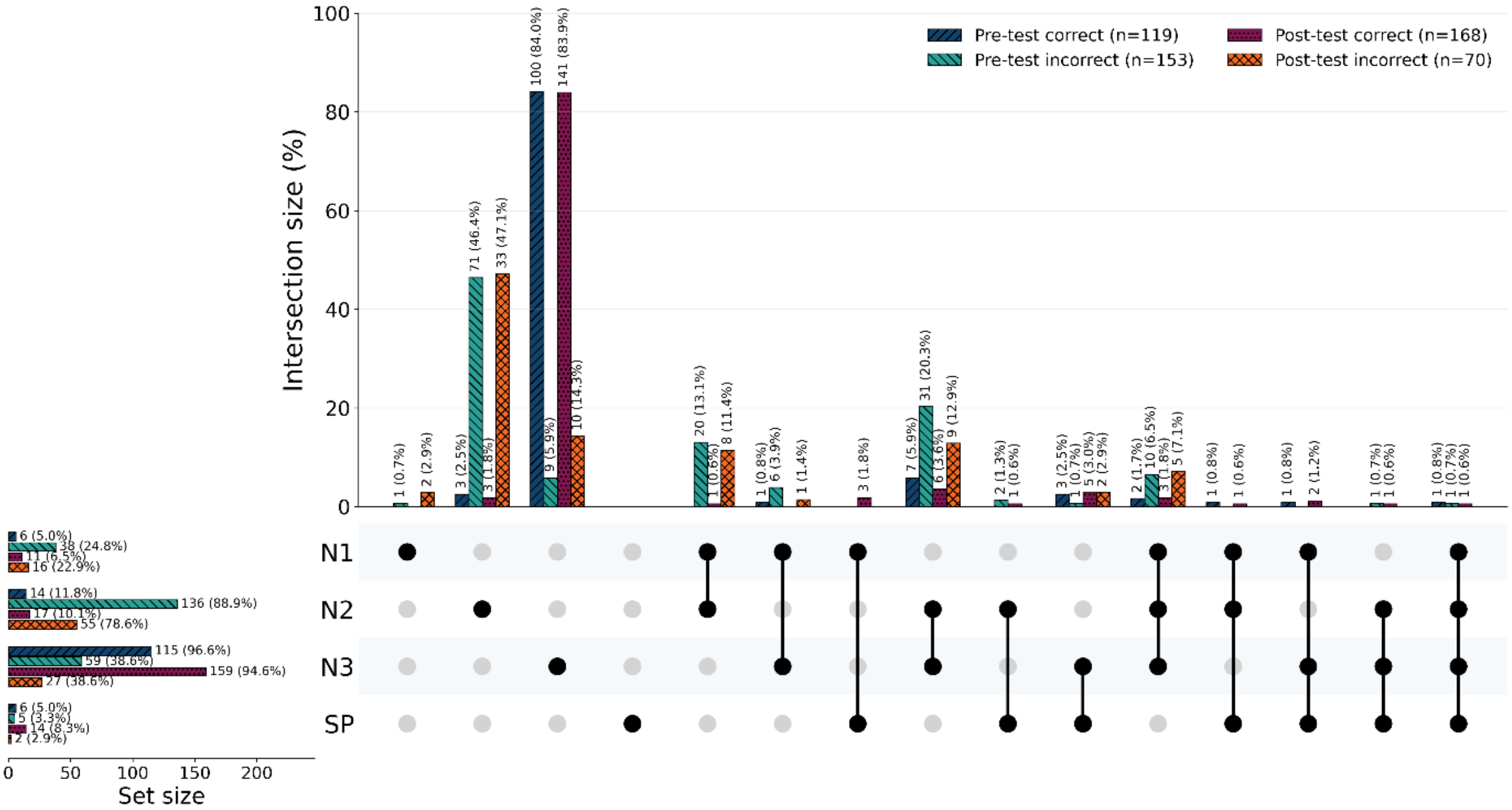

FIG. 5. UpSet plot showing responses to the law-selection sub-question Q15LS for students who gave the correct or incorrect response to Q15 both pre-test and post-test.

*2. Q16LS*

In contrast to the situation for Q15, most students gave the correct answer to Q16, both pre- and post-test. However, as shown in Fig 6, at pre-test, many students who gave the correct response to Q16 failed to select N3 in response to Q16LS, with only 35% of those who gave the correct response to Q16 selecting N3 alone, and 37% selecting N3 alongside other laws. Meanwhile a total of 50% of these students included N1 in their answer to Q16LS and a total of 31% included the Superposition Principle (SP) in their answer to Q16LS. Although the percentage giving the correct answer to Q16 only changed very slightly from pre-test to post-test, which could be attributed to the fact that not much learning was required, which is the premise of the ceiling effect, a different picture emerges when responses to Q16LS are considered. Many more of the students attributed their correct Q16 response to N3 by the time of the post-test, indicating that learning had taken place.

The number of incorrect responses to Q16 was so small (only 26 at pre-test and 18 at post-test) that it is not possible to interpret these students' responses to Q16LS in any meaningful way, so the responses have not been included here.

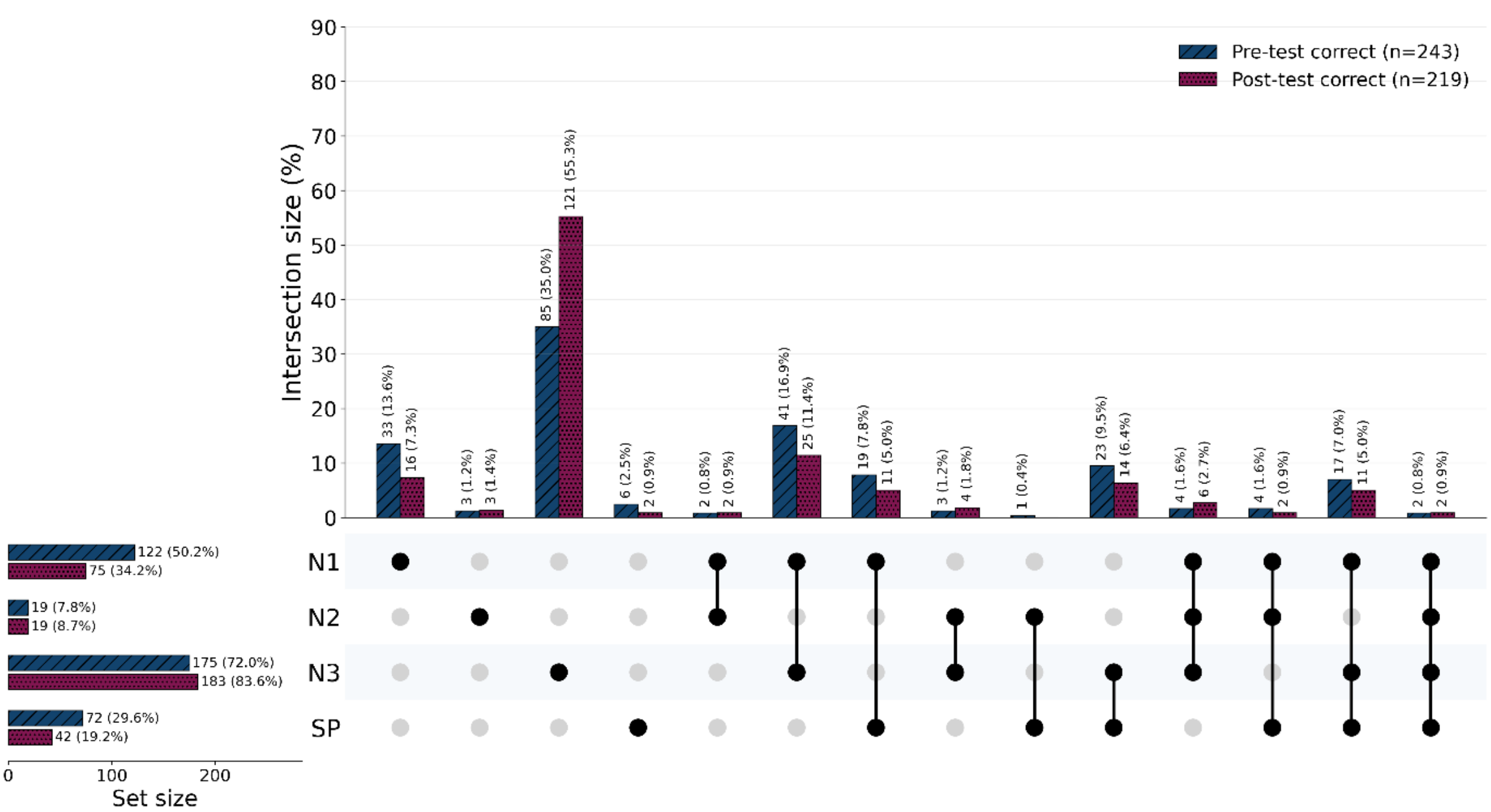

FIG. 6. UpSet plot showing responses to the law-selection sub-question Q16LS for students who gave the correct answer to Q16 pre-test and post-test.

### D. Longitudinal analysis of responses for selected students

Inspection of responses to Q15, Q15LS, Q16 and Q16LS from the same students reveals some interesting patterns, as illustrated in the examples given in Table V. Sometimes a student's responses to the law-selection questions confirms the nature of a student's conceptual understanding, as suggested by their responses to Q15 and Q16, and sometimes responses to Q15LS and Q15LS provide supplementary information. Finally, comparison of the responses to Q15 and Q15LS with the same student's responses to Q16 and Q16LS sometimes provides additional insight.

The pattern of responses to the four questions for the examples labelled as Student (1) and Student (2) reveal a good understanding of Newton's Third Law, while those for Student (3) reveal some misconceptions, even though the correct answer was given for Q16.

TABLE V. Responses given to Q15, Q15LS, Q16 and Q16LS by eleven example students.

| Student number | Response to Q15 (whether offered as MC or FT) | Response to Q15LS | Response to Q16 (whether offered as MC or FT) | Response to Q16LS |
|---|---|---|---|---|
| (1) | Option A (MC) | N3 | Option A (MC) | N3 |
| (2) | They are equal (FT) | N3 | They are equal (FT) | N3 |
| (3) | The car's force is greater (FT) | N2 | Option A (MC) | N1 |
| (4) | The force exerted by the car is greater as the car is accelerating (FT) | N2 | Option A (MC) | N1, SP |
| (5) | Force that the car exerts on truck is greater (FT) | N1, N2 | Forces are balanced (FT) | N1 |
| (6) | The force the car exerts on the truck is greater than that the truck exerts on the car (FT) | N1, N2 | They are now equal (FT) | N1 |
| (7) | It is equal (FT) | N3 | It is equal again (FT) | N3 |
| (8) | force that the car exerts is greater since there is a acceleration, while force that truck exerts is smaller (FT) | N2, N3 | the force that the car exerts is less than the force that the truck exerts since truck has higher mass (FT) | N1, N2 |
| (9) | They increase at the same rate (FT) | N3 | The forces are equal (FT) | N1, SP |
| (10) | Option A (MC) | N3 | Option A (MC) | SP |
| (11) | The are equal (Newton's third law) (FT) | N3 | They are still equal ("constant speed") (FT) | N1 |

Student (4)'s response to Q15 implies a use of Newton's Second Law in answering this question, which is supported by their response to Q15LS. Student (5)'s response to Q16 mentions "balanced", which was identified in the previous section as suggesting the use of Newton's First Law in answering the question, which is supported by their response to Q16LS. Similarly, the wording in Student (6) and Student (7)'s responses to Q16 (mention of "now equal" and "equal again" respectively), identified in the previous section as suggesting a poor or good Newtonian understanding, are in both cases supported by the student's response to Q16LS.

Student (9)'s answers to Q15 and Q16, which appear to show a confusion between force and acceleration and an attempt to the use Newton's Second Law in answering both questions, reveal more about their conceptual understanding than their responses to Q15LS and Q16LS. It is not clear why the student has selected N3 alongside N2 in Q15LS or selected N1 alongside N2 in Q16LS.

In other cases, a response which was classified as "X" but tentatively hypothesized to show some understanding (such as that given to Q15 by Student (9) was supported by the selection of N3 in the corresponding law-selection question (Q15LS in this case). However, after giving an apparently unambiguously correct answer to Q16, Student (9) then selected N1 rather than N3 for Q16LS. It was also observed that a number of students who had given apparently unambiguously correct answers to Q15 and selected N3 in response to Q15LS, then also correctly answered Q16 but selected N1 and/or SP in response to Q16LS, as illustrated by Student (10) and Student (11).

Quantitative investigation into this fascinating mismatch between students' responses to Q16LS and their earlier answers revealed that, at pre-test, of the 93 students who gave the correct answer to both Q15 and Q16, and selected N3 alone in their response to Q15LS, thus apparently illustrating robust

Newtonian understanding, 65 (70%) backed this up by selecting N3 alone in response to Q16LS. However, 9 of the students (10%) did not include N3 at all in their Q16LS response and 19 (20%) gave N3 alongside other options, most frequently the combination of N3 and SP (9 students).

At post-test, of the 129 students who gave the correct answer to both Q15 and Q16 and selected N3 alone in their response to Q15LS, 106 (82%) selected N3 alone in their response to Q16LS, with 17 (13%) selecting N3 along with other options (of which 7 selected N3 and N1) while just 6 (5%) did not include N3 at all in their response to Q16LS. Thus, not only did a higher percentage of students get the key questions correct, but those that did showed a more consistent conceptual understanding.

In a further investigation into correlations between student responses, of the total of 86 students who selected N3 alone in response to Q16LS at pre-test, 68 (79%) were found to have given the correct answer to Q15. In contrast, of the 84 students who did not include N3 at all in their response to Q16LS, just 17 (20%) gave the correct answer to Q15 and of the 99 students who included N3 in their answer to Q16LS along with other options, 31 (31%) gave the correct answer to Q15. There is thus a good correlation between correctness for Q16LS and correctness for Q15.

At post-test, improvements were again shown all round. 124 students now gave N3 alone in response to Q16LS and 108(87%) of these gave the correct answer to Q15. Of the 45 who omitted N3 in their response to Q16LS, 21(47%) gave the correct answer to Q15, while of the 68 who gave 103 along with other responses, 38 (56%) gave the correct answer to Q15.

**E. Comparison with data from other universities**

Fig. 7 and Fig. 8 compare the pre-test results for University A with the results from the combined dataset obtained from 112 students from the other five universities, for Q15 and Q16 respectively. These show that the MC and FT versions of both questions again give similar results, though students in the combined dataset were apparently even more successful at Q15 than was the case for University A pre-test, with Option A now being the most popular choice for both MC and FT format.

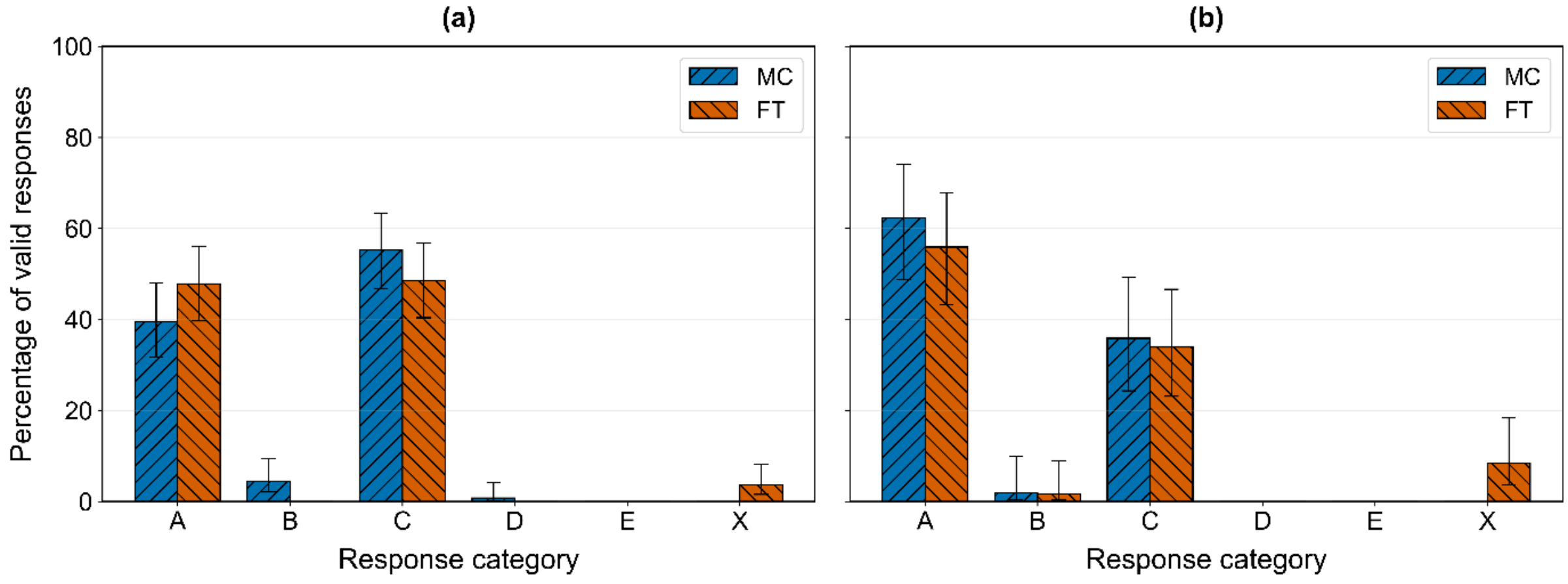

FIG. 7. Comparison of MC and FT pre-test responses to Q15 for (a) University A ($n$ = 272) and (b) Universities B-F combined ($n$ = 112). The error bars indicate 95% confidence intervals.

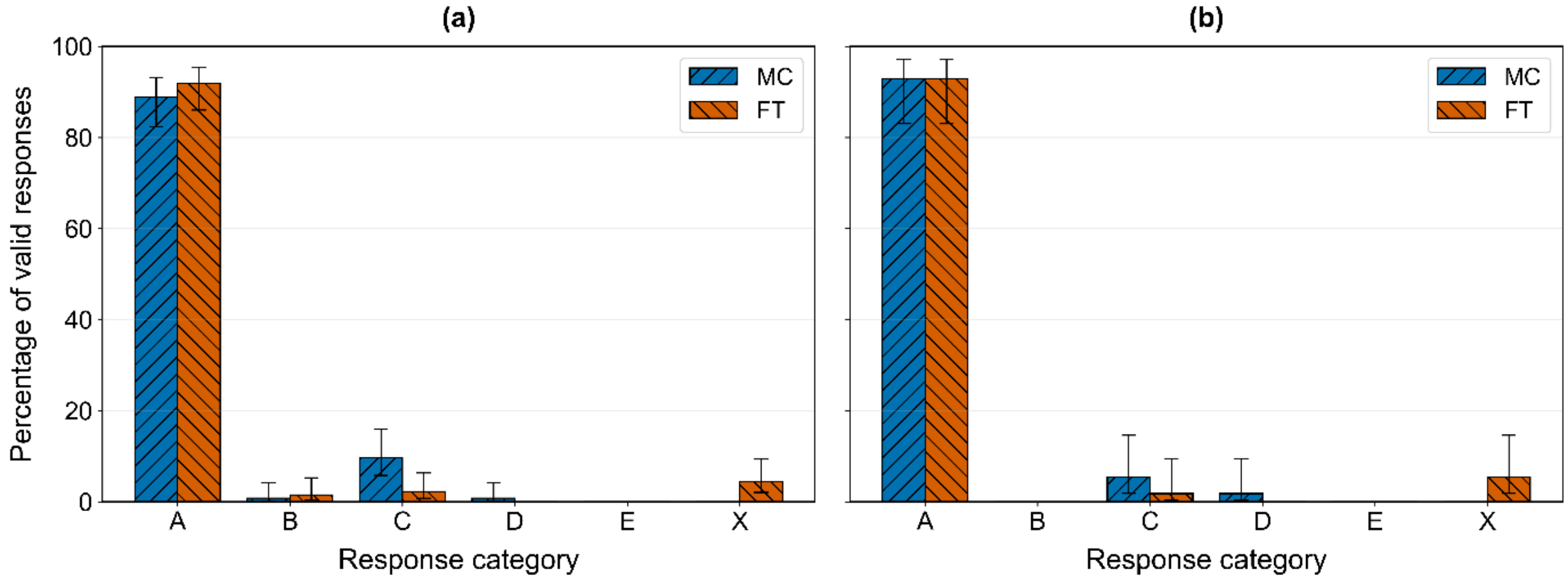

FIG. 8. Comparison of MC and FT pre-test responses to Q16 for (a) University A ($n$ = 269) and (b) Universities B-F combined ($n$ = 112). The error bars indicate 95% confidence intervals.

The results from the law-selection sub-questions are again in line with those for University A. Of those who gave the correct response to Q15, 97% selected N3 in response to Q15LS, while those who gave a different response to Q15, only 35% selected N3, while 80% selected N2 (Fig. 9) This finding becomes more pronounced when it is noted that the five students whose FT responses to Q15 were considered not to align with any of the MC options and thus classified at "X" all selected N3 alone in response to Q15LS. Their Q15 responses, given below, all included an element of truth, but they did not answer the question in the expected way:

- opposite forces
- they increase
- Both forces will increase equally
- they both increase in equal proportion
- Both increase in size, and are in opposite directions.

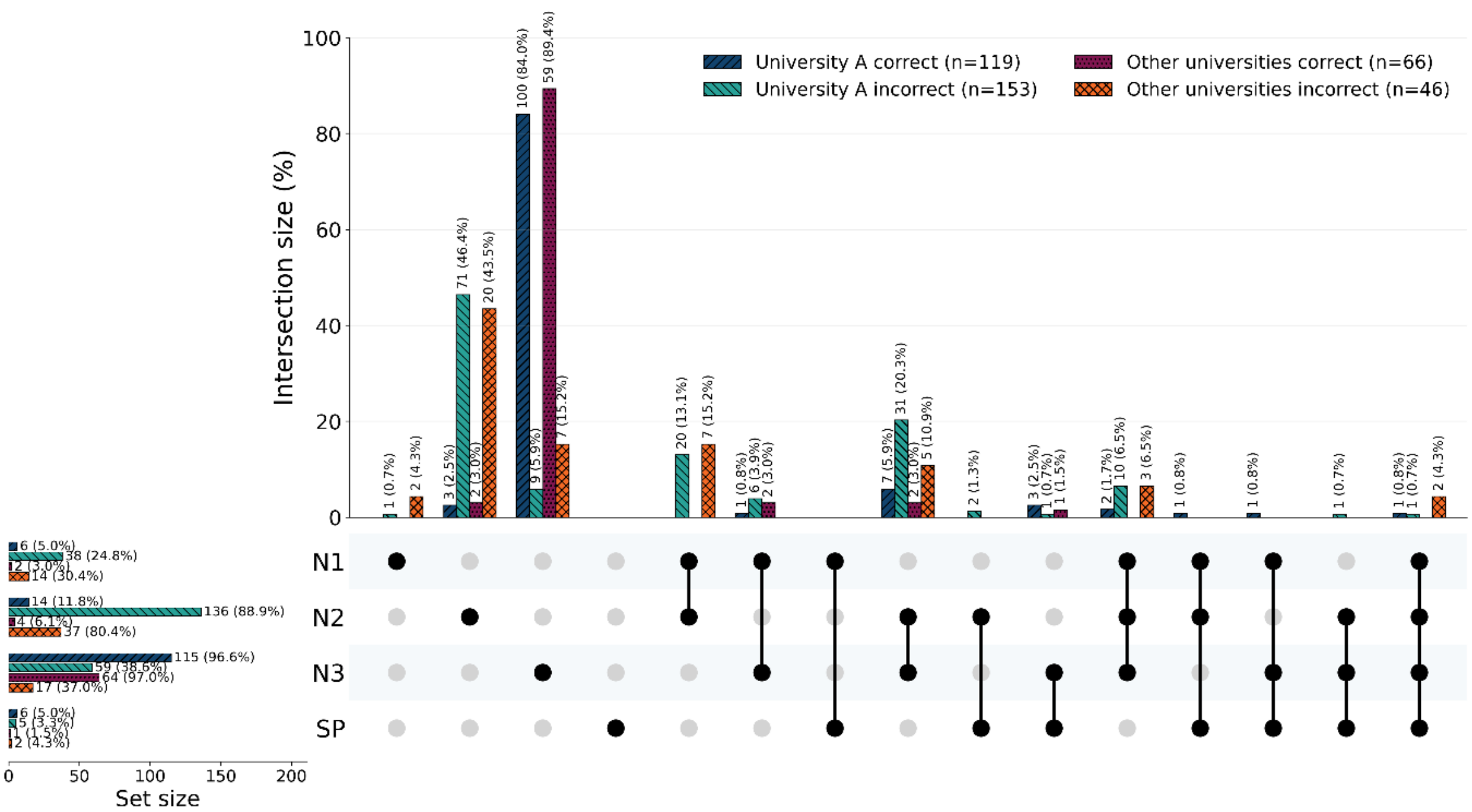

FIG. 9. UpSet plot showing responses to Q15LS for students from University A compared with Universities B-F who gave the correct answer and the incorrect answer to Q15.

104 (93%) of the students from Universities B-F gave the correct answer to Q16, but in a similar pattern to that observed for University A, responses to Q16LS revealed a less positive picture of conceptual understanding, as shown in Fig. 10. Although most selected N3, only 47% selected N3 alone, while 45% included N1 in their answer and 19% included the SP. Only eight students from Universities B-F gave incorrect responses to Q16 so it is impossible to draw generalizable conclusions from their responses to Q16LS. Three of these 8 Q16 responses were coded as X, and are given below, with the laws selected in Q16LS in brackets.

- Both forces will be 0 at there is no force exerted at constant velocity (N1)
- There is no resultant force so they balance each other out, but not necessarily in this scenario, as we do not know if we are suppose to take into account friction (N1)
- Car is exerting no force on truck therefore no force is exerted on the car from the truck (N1 and N3)

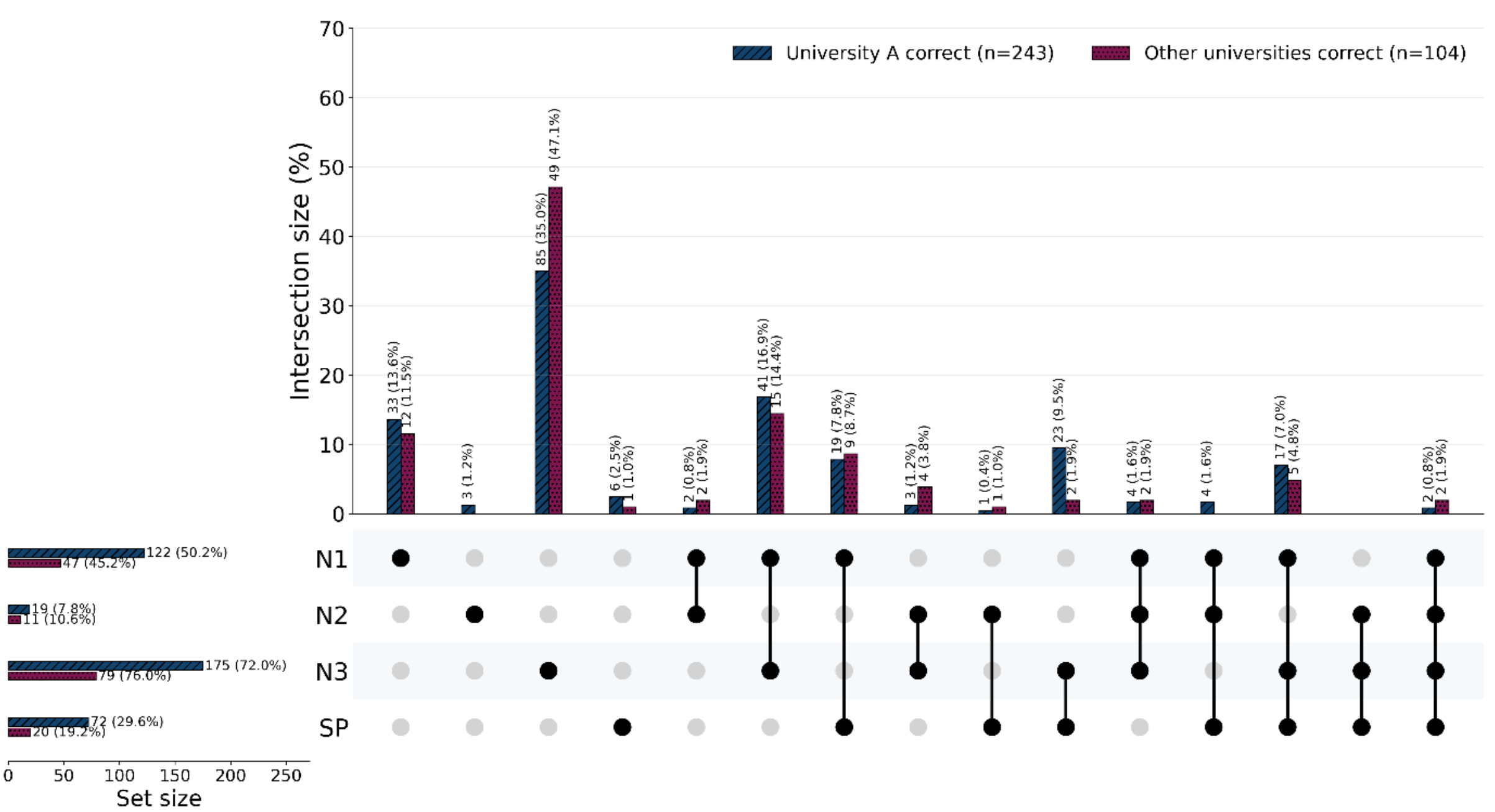

FIG. 10. UpSet plot showing responses to Q16LS for students from A compared with Universities B-F who gave the correct answer to Q16

Longitudinal review of responses to Q15, Q15LS, Q16 and Q16LS for Universities B-F show the same features as those observed for University A, as illustrated in the five examples given in Table VI.

Of the 58 students from Universities B-F who gave the correct answer to both Q15 and Q16, and selected N3 alone in their response to Q15LS, 36 (62%) then selected N3 alone in response to Q16LS. However, 8 of the students (14%) did not include N3 at all in their Q16LS response, with 7 students of these instead selecting N1. 14 (24%) gave N3 alongside other options, most commonly N3 and N1 (6 students).

Of the 49 students who selected N3 alone in response to Q16LS, 37 (76%) gave the correct answer to Q15. In contrast, of the 30 students who did not include N3 at all in their response to Q16LS, just 9 (30%) gave the correct answer to Q15. Of the 33 students who included N3 in their answer to Q16LS along with other options, 20 (61%) gave the correct answer to Q15.

TABLE VI. Responses given to Q15, Q15LS, Q16 and Q16LS by five example students from Universities B-F.

| Student number | Response to Q15 (whether offered as MC or FT) | Response to Q15LS | Response to Q16 (whether offered as MC or FT) | Response to Q16LS |
|---|---|---|---|---|
| (12) | Same (FT) | N3 | Same | N3 |
| (13) | The force the car exerts on the truck is larger so there is a resultant force forwards (FT) | N2 | Both forces are equal but act in opposite directions (FT) | N1 |
| (14) | Newton's third law states that every action has an equal and opposite reaction, so the forces should be equal (FT) | N3 | It stays the same | N3 |
| (15) | Both will increase in equal proportions (FT) | N3 | Option A (MC) | N3 |
| (16) | Option A (MC) | N3 | Option A (MC) | N1 |

The findings from Universities B-F relating to the performance of the FT questions and the law-selection questions in comparison to the conventional MC versions of FCI Q15 and Q16 thus showed consistently similar patterns to those for University A. However, the students in the combined dataset performed even better at pre-test than those from University A, especially on Q15. Their performance on the 14 questions from the Half FCI that were embedded within the MCQ was also high, with a mean Half FCI score of 10.68 ± 0.28 for the combined group of students from Universities B-F compared with 8.93 ± 0.21 for University A (pre-test).

## III. DISCUSSION

The baseline data from University A gathered in the five years prior to the main study reveal some significant differences from the data presented in 1992 in the paper that is commonly accepted as introducing the FCI [1], with significantly more students being successful in the University A data, especially pre-test, in both Q15 and Q16. The reason for this is unknown but possible factors include the different level of the students (and therefore different previous exposure to Newtonian Mechanics instruction), the different country and therefore educational system in which the work took place (UK vs US) and the fact that around 30 years had elapsed between the two studies; it is possible that changes in teaching methods have evolved in this time, leading to improved conceptual understanding. The potential impact of changing teaching methods and student experience was recognized back in 2004 [12] but the FCI has continued in widespread use.

It should also be noted that, early in the FCI's history, there were some minor changes in the wording of the Options presented to students (compare the version in [1] and the version in [2], which is the one commonly used, including at University A). Options A, B and C changed from “the amount of force of the car pushing against the truck” and “that of the truck pushing back against the car” to “The amount of force with which the car pushes on the truck” and “that with which the truck pushes back on the car”. Option B changed from one force being described as “less” than the other to being described as being “smaller” than the other. The original version of Option D was “the car’s engine is running so it applies a force as it pushes against the truck but the truck’s engine is not running so it can’t push back against the car, the truck is pushed forward simply because it is in the way of the car” but by 1995 this had been slightly simplified to “The car’s engine is running so the car pushes against the truck, but the truck’s engine is not running so the truck cannot push back against the car. The truck is pushed forward simply because it is in the way of the car.” It is difficult to see how the very small change in wording of Options A, B or C or even the slightly larger change in Option D might have caused the change in behavior, though the impact of seemingly innocuous changes of wording has been reported previously [31].

Option D and Option E both have two distinct aspects, for example “Neither the car nor the truck exerts any force on the other” and “The truck is pushed forward simply because it is in the way of the truck” for Option E. This complexity is recognized as a problem in the design of multiple-choice questions [32]. It takes more time for a student to read and interpret the complex options because they require the combining of multiple ideas that may be tightly linked for an expert in the subject but are far less closely linked for the student. Also, the student may consider part of the answer to be correct but not the other part. Several of the Q16 FT responses which we classified as "X" actually corresponded to the first part of Option E, but none corresponded to the whole statement given.

Comparison of the results for responses to FT and MC versions of the questions (Section B) with the earlier results from University A reveals more blank responses than in previous years. It is tempting to attribute this to a reluctance to type a free-text answer [33], but the effect is present in both formats, so it is most likely an artefact of the system used or some other unknown effect. Otherwise, the responses given to the FT versions of Q15 and Q16 show very similar results overall to those give to the MC versions, both in the current study and over the previous five years at University A. The conceptual understanding and misunderstanding revealed by the FT versions of the FCI questions was generally similar to that revealed by the MC versions. However, FT responses sometimes provided deeper insight into the student's conceptual understanding and into factors that should be considered when designing concept inventories.

Some of the FT responses did not answer the question in the intended manner (e.g. talking about both forces increasing rather than comparing them directly); particular care needs to be taken in the design of FT questions, where there are not predetermined options to scaffold a student's response. However, this lack of scaffolding also means that the student was not misled or inappropriately guided by the options, and subtleties of the student's conceptual model were sometimes revealed. Some of the FT responses received included a correct answer, but went on to add a qualification which revealed a misunderstanding, a surprisingly common feature of FT responses[34].

The law-selection sub-questions gave considerable additional insight into the conceptual understanding of the students in the study. Correct responses to Q15 were usually supported by the selection of Newton's Third Law in Q15LS. In the case of incorrect responses to Q15 (almost always Option C), most of the FT responses that revealed more of a student's understanding mentioned "acceleration", giving some indication that the student was using Newton's Second Law reasoning, and answers to Q15LS supported this.

The most startling finding of the current study was the large number of students whose correct response to Q16 would, by itself, indicate good conceptual understanding, but they then justified this response (sometimes explicitly by a quantification in Q16 and more commonly in Q16LS) by indicating that they had used Newton's First Law or the Superposition Principle in reaching their answer. Rather than using reasoning based on an equal-and-opposite interaction pair (a Third Law relationship between the two objects) the students were confusing two balanced forces on a single object. This source of confusion was recognized in 1992, but for the intervening 34 years, it has been tempting to assume that students who have given the "correct" answer to Q16 have good conceptual understanding, which is not the case. It is thus not surprising that so many more students give the correct answer to Q16 than give the correct answer to Q15.

Furthermore, while the high apparent pre-test success rate for Q16 would, by itself, imply that there was little room for improvement, the marked improvement in performance on Q16LS from pre-test to post-test indicates that instruction was effective, a learning gain that would be hidden in results based on the conventional FCI.

It is interesting that a significant number of students gave the same response to Q15 and Q16 but different answers to Q15LS and Q16LS.  This result casts some doubt over these students' conceptual understanding, but needs further investigation, perhaps by follow-up interviews after a repeat of the current study. Interviews would allow further exploration of unclear free-test responses.

Another area of work for the future would be a detailed investigation into differences in responses from male and female students. Preliminary work [35] based on the current dataset has shown a gender gap, with students who self-declare to be male achieving better results for Q15 than those who self-declare to be female, but the gap is smaller for Q16, as expected by the high percentage of all students who give the correct answer for this question. There are some tantalizing indications of a more complex pattern when results for the law selection questions are considered.

Most of the results presented here have been from students at a single UK university, so further investigation at a range of institutions would also be beneficial. The generally similar results from Universities B-F gives an early indication that the results presented have widespread applicability, though the reasons for the high pre-test performance from students in this group is not known. One possible explanation is that more than half the students in the combined group were from University B, most of whom will have obtained the top A* grade in physics A-level, the most common school leaving qualification in England (although other entry routes are possible). Students will have had a thorough grounding in Newtonian mechanics as part of this qualification. Another possible explanation is that only the students from University A completed the NMQ under invigilation; those from Universities B-F completed the NMQ in their own time and they may have had assistance from Generative Artificial Intelligence.

The current version of the FCI uses the same options for Q15 and Q16, but if the results we present are replicated elsewhere, a relatively straightforward improvement would be to amend the distractors to bring them into line with the responses revealed by free-test and law-selection questions. More generally, we encourage the use of concept inventory questions that delve more deeply into conceptual understanding. The rise of Generative Artificial Intelligence presents a significant challenge to many forms of assessment [36], but it also accelerates the potential for the automatic marking of free-text responses [37], including a potential to map from longer answers to the concepts the student has used [23].

This article has presented detailed findings from four of the questions in a modified Force Concept Inventory. Wider findings about the use of alternative question types and student understanding of Newtonian mechanics and are presented in the doctoral thesis on which the article is based. Additional insights are likely to be available from detailed analysis of the data [35] for other sub-questions.

## IV. CONCLUSION

Data from six UK universities has shown a similar pattern of results for multiple-choice and free-text versions of two FCI questions, also comparable with conventional FCI results from one of the universities in the preceding five years, but with marked differences relative to early published FCI data. Free-text responses and law-selection sub-questions revealed more about conceptual understanding than could be found from multiple-choice responses alone.

In particular, when comparing the force from a car on a truck it is pushing with the force from the truck on the car, when the vehicles are moving at constant speed, many students were found to confuse the correct equal-and-opposite interaction pair between two objects with the balanced forces on a single object moving at constant speed. This reveals a fundamental misunderstanding of Newton's Third Law which was not apparent from the FCI alone. The observed improvement in post-instruction responses to the law-selection sub-question revealed a learning gain that would have been hidden in conventional FCI results.

The use of novel question types presents an opportunity to revolutionize concept inventories and their use in STEM education in the future.

ACKNOWLEDGEMENTS

The authors thank the students who participated in pilot testing and interviews as part of development of the Newtonian Mechanics Quiz and those who took part in the study described here. We also thank the many colleagues who have given willingly of their time to advise, formally and informally, on the development of the Newtonian Mechanics Quiz.

DATA AVAILABILITY

The anonymized data from the Newtonian Mechanics Quiz, supporting the findings of this article are publicly available via the UK Open University's Open Research Data Online repository [38].

The study was approved by the UK Open University's Human Ethics Research Committee, reference 2024-0387-3.

APPENDICES

Appendix 1: Table VI presents historical data for Q15 of the FCI from University A for academic years from 2019-20 to 2023-24. Pre-test and post-test results show little variation between years and can therefore be combined for comparison with other administrations of the FCI. Table VII presents similar results for FCI Q16.

TABLE VI. Detailed results from University A for Q15 for the academic years 2019-20 to 2023-24.

| **Academic year** | **Pre- or Post-test** | **Number of students** | **Number (percentage) of students who chose option** | | | | | **Number (%) of blank responses** |
|---|---|---|---|---|---|---|---|---|
| | | | **A** | **B** | **C** | **D** | **E** | |
| 2019-20 | Pre-test | 155 | 56 (36%) | 3 (2%) | 94 (61%) | 0 (0%) | 1 (1%) | 1 (1%) |
| | Post-test | 125 | 90 (72%) | 3 (2%) | 31 (25%) | 0 (0%) | 0 (0%) | 1 (1%) |
| 2020-21 | Pre-test | 214 | 100 (47%) | 4 (2%) | 103 (48%) | 1 (<1%) | 0 (0%) | 6 (3%) |
| | Post-test | 136 | 98 (72%) | 0 (0%) | 35 (26%) | 1 (1%) | 0 (0%) | 2 (1%) |
| 2021-22 | Pre-test | 240 | 89 (37%) | 6 (3%) | 138 (58%) | 3 (1%) | 1 (<1%) | 3 (1%) |
| | Post-test | 163 | 100 (61%) | 2 (1%) | 59 (36%) | 0 (0%) | 0 (0%) | 2 (1%) |
| 2022-23 | Pre-test | 211 | 87 (41%) | 3 (1%) | 118 (56%) | 1 (<1%) | 0 (0%) | 2 (1%) |
| | Post-test | 102 | 70 (69%) | 1 (1%) | 31 (30%) | 0 (0%) | 0 (0%) | 0 (0%) |
| 2023-24 | Pre-test | 278 | 109 (39%) | 6 (2%) | 159 (57%) | 4 (1%) | 0 (0%) | 0 (0%) |
| | Post-test | 201 | 157 (78%) | 2 (1%) | 40 (20%) | 2 (1%) | 0 (0%) | 0 (0%) |
| 2019-24 combined | Pre-test | 1098 | 441 (40%) | 22 (2%) | 612 (56%) | 9 (1%) | 2 (<1%) | 12 (1%) |
| | Post-test | 727 | 515 (71%) | 8 (1%) | 196 (27%) | 3 (<1%) | 0 (0%) | 5 (1%) |

TABLE VII. Detailed results from University A for Q16 for the academic years 2020-21 to 2023-24.

| Academic year | Pre- or Post-test | Number of students | Number (percentage) of students who chose option A | B | C | D | E | Number (%) of blank responses |
|---|---|---|---|---|---|---|---|---|
| 2020-21 | Pre-test | 134 | 124 (93%) | 0 (0%) | 5 (4%) | 1 (1%) | 0 (0%) | 4 (3%) |
| | Post-test | 84 | 77 (92%) | 0 (0%) | 4 (5%) | 0 (0%) | 1 (1%) | 2 (2%) |
| 2021-22 | Pre-test | 219 | 201 92%) | 1 (0%) | 8 (4%) | 3 (1%) | 3 (1%) | 3 (1%) |
| | Post-test | 153 | 144 (94%) | 1 (1%) | 1 (1%) | 0 (0%) | 5 (3%) | 2 (1%) |
| 2022-23 | Pre-test | 200 | 182 (91%) | 1 (1%) | 11 (6%) | 0 (0%) | 4 (2%) | 2 (1%) |
| | Post-test | 99 | 96 (97%) | 0 (0%) | 2 (2%) | 0 (0%) | 0 (0%) | 1 (1%) |
| 2023-24 | Pre-test | 257 | 232 (90%) | 1 (<1%) | 16 (6%) | 0 (0%) | 8 (3%) | 0 (0%) |
| | Post-test | 190 | 184 (97%) | 2 (1%) | 2 (1%) | 1 (1%) | 1 (1%) | 0 (0%) |
| 2019-24 combined | Pre-test | 810 | 739 (91%) | 3 (<1%) | 40 (5%) | 4 (<1%) | 15 (2%) | 9 (1%) |
| | Post-test | 526 | 501 (95%) | 3 (1%) | 9 (2%) | 1 (<1%) | 7 (1%) | 5 (1%) |